\documentclass[twocolumn,showpacs,preprintnumbers,amsmath,amssymb,apl]{revtex4}

\usepackage{graphicx}
\usepackage{dcolumn}
\usepackage{bm}
\input{epsf}

\begin{document}

\newcommand{\note}[1]{\textbf{#1}}

\title{
Dispersive readout scheme for a Josephson phase qubit }
\author{T.~Wirth, J.~Lisenfeld, A. Lukashenko, and A.~V.~Ustinov}

\email{ustinov@kit.edu}

\affiliation{Physikalisches Institut, Karlsruhe Institute of Technology and\\
DFG-Center for Functional Nanostructures (CFN)\\
D-76128 Karlsruhe, Germany}

\date{\today}

\begin{abstract}
We present experimental results on a dispersive scheme for reading
out a Josephson phase qubit. A capacitively shunted dc-SQUID is used
as a nonlinear resonator which is inductively coupled to the qubit.
We detect the flux state of the qubit by measuring the amplitude and phase of a microwave pulse reflected from the SQUID resonator. By this low-dissipative method, we reduce the qubit state measurement time down to
25 $\mu$s, which is much faster than using the conventional readout
performed by switching the SQUID to its non-zero dc voltage state.
The demonstrated readout scheme allows for reading out multiple
qubits using a single microwave line by employing frequency-division multiplexing.
\end{abstract}



\maketitle

A vital ingredient of experiments on quantum bits is a detection
tool to efficiently read out the state of a qubit. This detector
must introduce as little back-action as possible, while showing a
large measurement contrast; it should have negligible dissipation
and offer fast operation. Superconducting quantum bits, such as the
flux~\cite{Mooij-Sci-99} and phase qubit~\cite{Martinis-PRL-02},
consist of superconducting loops interrupted by one or more
Josephson junctions. Since their readable states can be
discriminated by the magnetic flux passing through the qubit loop,
it is common to use inductively coupled dc-SQUIDs as sensitive
detectors.

The standard method to read out a Josephson phase qubit is to record
the dc bias current at which the SQUID switches to its
non-superconducting state~\cite{Martinis-PRL-04, Wellstood-PRB-06,
Buisson-PRB-07, Lisenfeld-PRL-07}. This process generates heat
directly on the chip and quasi-particles in the circuitry. Both
effects are responsible for a relatively long cool-down time of
about 1-2 ms that is required after each switching event. This,
together with the time needed to ramp up the bias current of the
SQUID, limits the repetition rate of the experiment.

For the flux qubit, non-destructive dispersive readout schemes have
successfully been realized already some time ago, either by coupling
to a high quality LC-tank circuit~\cite{Ilichev-PRL-03}, or to a
dc-SQUID~\cite{Lupascu-PRL-04}.
So far, most measurements of phase qubits were typically done
by the above mentioned switching current measurement of an
inductively coupled dc-SQUID. Recently, first experiment overcoming
the limitations of the switching readout was
reported~\cite{IBM-APL-10}. In this approach, the phase qubit was
capacitively coupled to a transmission line which allows for direct
probing its resonance frequency with a microwave pulse. This
approach eliminates a readout dc-SQUID, but in turn introduce
decoherence via the line coupled directly to the qubit.

In this letter, we present experiments on dispersive readout of a
SQUID weakly coupled to a phase qubit. By using weak
coupling between the SQUID and the qubit, this scheme
protects the qubit from decoherence
sources introduced by the readout circuitry. Moreover, while
preserving the intrinsic coherence of the qubit, this method is
suitable for reading out many qubits using a single microwave line and
frequency-division multiplexing addressing individual readout SQUID
resonators.

\begin{figure}[!bht]
    \includegraphics[width=0.7\columnwidth]{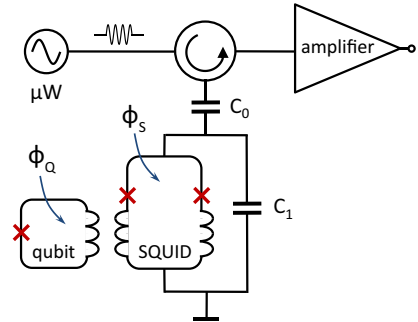}
    \caption{(Color online) Scheme of the measurement setup. The SQUID with shunt capacitor $\mathrm{C_\mathrm{1}}$ coupled to the qubit. The pulsed microwave signal is applied via a cryogenic circulator, and the reflected signal is amplified by a cryogenic amplifier.}
    \label{fig:1}
\end{figure}

We couple the qubit to a capacitively shunted dc-SQUID which forms a
tank circuit having a resonance frequency around 2 GHz. It is
connected to a microwave line by a coupling capacitor
$\mathrm{C_{0}}$ shown in Fig.~\ref{fig:1}. Our sample was
fabricated in a standard niobium-aluminium trilayer process.
Measurement of the amplitude and phase of a reflected microwave
pulse allows one to determine the shift of the resonance frequency
of the SQUID-resonator and by doing this deduce the magnetic flux of
the qubit state.

Depending on the applied microwave power, the SQUID resonator
circuit can be operated in either linear or nonlinear regime. The
nonlinear regime makes it possible using extremely sensitive bifurcation
readout~\cite{Siddiqi-RevSciInstr-09}. This may allow for a direct quantum
non-demolition readout of a Josephson phase qubit, similar to what
has already been realized for a flux
qubit~\cite{Lupascu-NatPhys-07}.

Figure~\ref{fig:1} shows a simplified scheme of our measurement
setting. At room temperature, the output of a continuous-wave
microwave source is split using a power divider into reference and
probe signals (not shown). The probe signal passes through a
phase shifter and is amplitude-modulated by means of two mixers
connected in series in order to achieve a large on-off ratio of
about 64 dB. The pulsed signal is then attenuated in total by 70 dB
using attenuators at several temperature stages of the dilution
refrigerator. At the 30 mK stage, the signal is passed through a
cryogenic circulator (Pamtech STE1438K) with isolation of 18 dB in
the frequency range between 1.9 GHz and 2.4 GHz. From there, the
signal is guided to the SQUID circuit by an on-chip coplanar
transmission line.

The signal reflected from the sample goes back to the circulator
towards a cold amplifier installed at the 4 K stage, specified to 42
dB gain at a noise temperature of 6 K (Quinstar QCA-S.3-30H). The
narrow-band isolator together with a filter (Mini-Circuits
VBFZ-2130) of pass band between 1.7 GHz and 2.4 GHz serve to protect
the sample from the noise of the amplifier. At the output of
the amplifier, a low-pass filter (Mini-Circuits VLF-3000) with a
cut-off frequency of 3 GHz protects the amplifier and the whole
system from high frequency noise. At room temperature, the signal is
amplified by two amplifiers (Mini-Circuits ZX60-2534M) of total gain
76 dB and mixed together with the reference signal in an IQ mixer
(Marki IQ1545LMP). The $I$ and $Q$ outputs of this mixer, which are
determined by a combination of the phase and amplitude of the
signal, are digitized by means of an 8-bit, 100 MS/s data
acquisition card.

On chip, there are two magnetic flux lines, one for flux
$\Phi_\mathrm{S}$ biasing the SQUID, see Fig.~\ref{fig:1}, and
another for flux $\Phi_\mathrm{Q}$ biasing the qubit. The qubit is
controlled by microwave pulses which are applied via a separate line
(not shown) attenuated at several low temperature stages. The SQUID
flux bias line is equipped with a current divider and filter at the
1 K stage, and a powder filter~\cite{Lukash-RevSciInstr-08} at the
sample holder. By taking the crosstalk of the two flux coils into
account, we can independently change the flux that is seen by the
qubit and the flux that is seen by the SQUID. This sample was
designed with a large mutual inductance between qubit loop and
dc-SQUID which allowed us to independently characterize the sample
by the conventional switching-current technique. The dispersive
readout results presented below were obtained without applying any
dc-bias to the readout SQUID.

To operate the phase qubit, we biased it close to one flux
quantum in its loop, giving rise to an asymmetric double-well
potential~\cite{Martinis-PRL-04}. The qubit quantum states are
located in the shallow metastable well and can be distinguished by
their tunneling rate to the neighboring deep well. In order to read
out the qubit, this tunneling is triggered by applying a short
(about 1 ns) flux pulse of small amplitude which tilts the potential
such that tunneling occurs nearly exclusively only from the excited
state. Since states in neighboring wells differ by the number of
flux quanta in the loop, reading out the qubit is completed by
measuring the resulting magnetic field via the inductively coupled
dc-SQUID.

\begin{figure}[!bht]
    \includegraphics[width=\columnwidth]{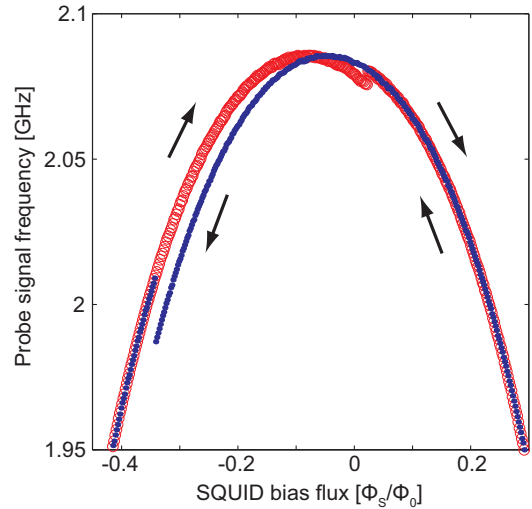}
    \caption{(Color online) Microwave frequency applied to the SQUID vs. externally applied flux. The measurement points show the position of a dip in the reflected signal amplitude for two different directions of the flux sweep.}
    \label{fig:2}
\end{figure}

The  the position of a dip in the amplitude of the reflected pulse is plotted in
Fig.~\ref{fig:2} as a function of microwave frequency and
applied SQUID flux bias $\Phi_\mathrm{S}$. Data points indicate the
dependence of the tank circuit resonance frequency on the applied bias flux. The larger (red)
circles correspond to the flux swept from negative to positive
values, while the smaller (blue) dots stand for the flux swept in
opposite direction. During the flux sweep, due to the crosstalk
between $\Phi_\mathrm{S}$ and $\Phi_\mathrm{Q}$ flux lines
approximately one flux quantum $\Phi_0$ enters or leaves the qubit
loop, which gives rise to abrupt shift of the dip frequency at specific flux bias
values. The resonance frequency shift at a bias flux of -0.35 $\Phi_0$ is about 22 MHz, which is larger than the tank
circuit's resonance line width of about 4 MHz. This scheme is thus capable
of single-shot detection of the qubit flux state.

\begin{figure}[!bht]
    \includegraphics[width=\columnwidth]{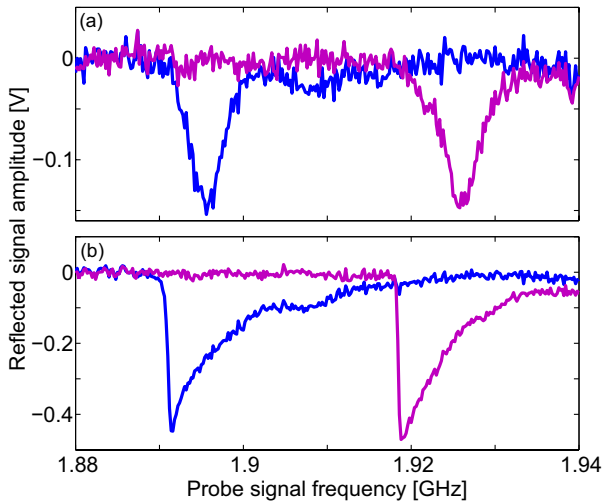}
    \caption{(Color online) Shift of the resonance frequency of the SQUID resonator by 30 MHz due to the qubit changing its magnetic flux by approximately $\Phi_0$. (a) In the linear regime. (b) SQUID driven in the non-linear regime. Note the larger signal amplitude compared to the linear regime.}
    \label{fig:3}
\end{figure}

The SQUID resonator frequency shift induced by the qubit is shown in detail in Fig.~\ref{fig:3}(a). It displays two traces of the normalized
reflected signal amplitude versus the applied microwave frequency in
the vicinity of the qubit-state switching. Here, the resonance was located at around 1.9 GHz where the SQUID has higher sensitivity to the flux. The amplitude of the reflected signal drops at the resonance frequency. For this measurement, very low microwave power of -120 dBm was applied to SQUID to stay in the linear regime, giving rise to the Lorentzian shape of the
resonance dips. Taking into account the line width of 4 MHz and
the dependence of the resonance frequency on the flux, we achieve a
flux resolution of 2-3 m$\Phi_0$ of the detector at operating
frequency of 1.9 GHz. As the two qubit states differ by magnetic
flux of the order of $\Phi_0$, this allows for a very weak inductive
coupling between SQUID and qubit for future experiments.
Fig.~\ref{fig:3}~(b) shows the same frequency range as above, but
now the power of the input signal is larger, -115 dBm, driving the
SQUID into the nonlinear regime. This is revealed by the shape of the
dips. The advantage of the non-linear regime is the sharper edge on
the low frequency side which allows for an even better flux
resolution of about 0.5-0.7 m$\Phi_0$.

\begin{figure}[!bht]
    \includegraphics[width=\columnwidth]{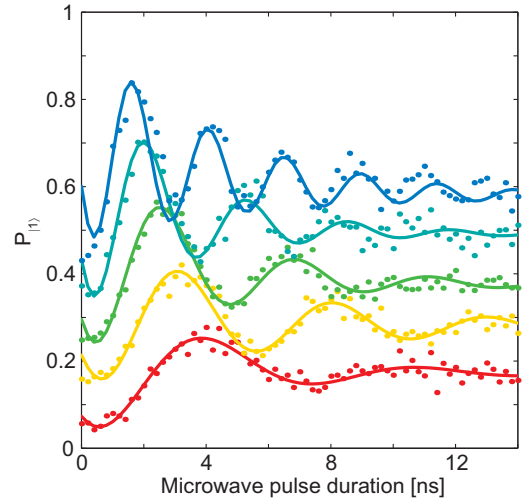}
    \caption{(Color online) Coherent oscillations of the qubit for different driving powers, from bottom to top: -18 dBm, -15 dBm, -12 dBm, -9 dBm and -6 dBm. Curves are offset by 0.1 for better visibility.}
    \label{fig:4}
\end{figure}

Figure~\ref{fig:4} shows Rabi oscillations of the qubit measured
for different driving powers of the qubit microwave driving. As it is expected, the frequency of Rabi oscillations increases approximately linearly with the driving field amplitude. The measured energy relaxation time of the tested qubit is rather short and is of order of $T_{1}$ = 5 ns. This time is it not limited by the chosen type of readout but rather determined by the intrinsic coherence of the qubit itself. We verified this fact by measuring the same
qubit with the conventional SQUID switching current method, which yielded very similar $T_{1}$. The observed short coherence time is likely to be caused by the dielectric loss in the silicon oxide forming the insulating dielectric layer around the qubit Josephson junction~\cite{Lisenfeld-PRL-07}.

In conclusion, we have demonstrated a dispersive readout scheme for a
Josephson phase qubit. This scheme avoids the switching of the SQUID flux detector into a resistive state. Due to much lower dissipation in the circuit, we we can reduce SQUID measurement time down to 25 $\mu$s without observing any noticeable heating effects. This readout repetition time is
about 40 times shorter than the time typically achieved with the conventional readout. In our setup, the shortest repetition time is limited by the amplifiers for the $I$ and $Q$ signals and could be further reduced down to the 10~-~100 nanosecond range by expanding the band of these amplifiers,
as it has been already demonstrated for flux
qubits~\cite{Lupascu-NatPhys-07}. This short measurement time and
the possibility of using frequency-division multiplexing
readout~\cite{Irwin-APL-08} make our approach promising for future
experiments scaled up to multiple phase qubits.

We acknowledge financial support from the EU project SOLID and
the Deutsche Forschungsgemeinschaft (DFG).

\end{document}